# Radiative neutron capture on $^{16}$O at astrophysical energies

S B Dubovichenko[1,2], N V Afanasyeva[1,3] and A V Dzhazairov-Kakhramanov[1,2]

[1] V. G. Fessenkov Astrophysical Institute "NCSRT" NSA RK, 050020, Observatory 23, Kamenskoe plato, Almaty, Kazakhstan

[2] Institute of nuclear physics NNC RK, 050032, str. Ibragimova 1, Almaty, Kazakhstan

[3] Al-Farabi Kazakh National University, 050040, av. Al-Farabi 71, Almaty, Kazakhstan

E-mails: dubovichenko@mail.ru; n.v.afanasyeva@gmail.com; albert-j@yandex.ru

**Abstract**
The total cross sections of the $^{16}$O(n, $\gamma$)$^{17}$O radiative capture process is described in the framework of the modified potential cluster model with the classification of orbital states according to Young tableaux. The possibility to describe the experimental data for the total cross sections of the neutron radiative capture on $^{16}$O at the energies from 0.025 eV to 1.5 MeV is shown on the basis of $E$1 and $M$1 processes. The transitions to the bound states of $^{17}$O in the n$^{16}$O channel from $^2S_{1/2}$, $^2P_{3/2}$ and $^2D_{3/2}$ waves of the n$^{16}$O scattering have been considered. These transitions have been considered in the resonance energy region at 433 and 1000 keV.

## 1. Introduction

Light radioactive nuclei play an important role in many astrophysical environments. Such parameter as cross section of the capture reactions as a function of energy is very important for investigation of many astrophysical phenomena such as primordial nucleosynthesis of the Universe, main trends of stellar evolution, novae and super-novae explosions, X-ray bursts etc. The continued interest in the study of processes of radiative neutron capture on light nuclei at low and ultralow energy is caused by several reasons. Firstly, this process plays a significant part in the study of many fundamental properties of nuclear reactions, and secondly, the data on the capture cross sections are widely used in a various applications of nuclear physics and nuclear astrophysics, for example, in the process of studying of the primordial nucleosynthesis reactions. Therefore, continuing the analysis of the processes of radiative capture of neutrons by light atomic nuclei, which are part of the different thermonuclear cycles [1,2,3,4] and in one of the variants of reaction chains for inhomogeneous Big Bang models [5-9],

$$^{1}\text{H}(n, \gamma)^{2}\text{H}(n, \gamma)^{3}\text{H}(^{2}\text{H}, n)^{4}\text{He}(^{3}\text{H}, \gamma)^{7}\text{Li}(n, \gamma)^{8}\text{Li}(^{4}\text{He}, n)^{11}\text{B}(n, \gamma)^{12}\text{B}(\beta^{-})$$
$$^{12}\text{C}(n, \gamma)^{13}\text{C}(n, \gamma)^{14}\text{C}(p, \gamma)^{15}\text{N}(n, \gamma)^{16}\text{N}(\beta^{-})\underline{^{16}\text{O}(n, \gamma)^{17}\text{O}}\ldots . \qquad (1)$$

Let us consider the n$^{16}$O → $^{17}$Oγ capture reaction at astrophysical energies, which apparently flowed at different stages of the creation, development and the formation of the Universe. This reaction is also important for *s*-processes for stars of various metallicity. [1,2,10,11].

The extremely successful development of microscopic models like Resonating Group Method (RGM, see, for example, [12,13]), Generator Coordinate Method (GCM, see, particularly, [14]) or algebraic version of RGM [15], leads to the view that the advancement of new results in low energy nuclear physics and nuclear astrophysics is possible only in this direction. Eventually, very common opinion states that this is the only way in which the future development of our ideas about structure of atomic nucleus, nuclear and thermonuclear reactions at low and astrophysical energies can be imagined.

However, the possibilities of simple two-body potential cluster model (PCM) were not completely studied up to now, particularly, if it use the conception of forbidden states (FS) [16] and directly take into account the resonance behavior of elastic scattering phase shifts of interacting particles at low energies [10]. This model can be named as modified PCM (MPCM with FS or simply MPCM). The potentials of the bound states (BS) in this model are constructed on the basis of the description not only the phase shift, binding energy and charge radii, but also the asymptotic constants (AC) in the specified channel. The rather difficult RGM calculations are not the only way to explain the available experimental facts. Simpler MPCM with FS can be used, taking into account the classification of orbital states according to Young tableaux and the resonance behavior of the elastic scattering phase shifts. Such approach, as it was shown earlier [10,11,16], in many cases allows one to obtain quite adequate results in description of many experimental data.

The results of the phase shift analysis of the experimental data on the differential cross sections for the elastic scattering [17] of corresponding free nuclei correlated to such clusters [18-21] are usually used for the construction of the interaction potentials between the clusters for the scattering states in the MPCM. Intercluster interaction potentials within the framework of the two-particle scattering problem are constructed from the best description of the elastic scattering phases [21-23]. Moreover, for any nucleon system, the many-body problem is taken into account by the division of single-particle levels of such potential on the allowed and the forbidden by the Pauli principle states [18-21]. However, the results of phase shift analysis, being usually available only in a limited range of energies and with large errors, do not allow, as a rule, to reconstruct completely uniquely the interaction potential of the scattering processes and the BSs of clusters. Therefore, an additional condition for the construction of the BS potential is a requirement of reproduction of the nucleus binding energy in the corresponding cluster channel and a description of some other static nuclear characteristics, such as the charge radius and the AC.

This additional requirement is, of course, an idealization of real situations in the nucleus, because it suggests that there is 100% clusterization in the ground state. Therefore, the success of this potential model to describe the system of *A* nucleons in a bound state is determined by how high the real clusterization of this nucleus in the $A_1 + A_2$ nucleons channel. At the same time, some of the nuclear characteristics of the individual, not even the cluster, nuclei can be mainly caused by one particular cluster channel. In this case, the single-channel cluster model used here allows us to identify the dominant cluster channel, to identify and describe those properties of nuclear systems that are caused by it [18-20]. In addition, the potentials of the BS and the continuous spectrum, taking into account the resonant nature of the elastic scattering phase shifts in a given partial wave must meet the classification of cluster orbital states according to the Young tableaux.

## 2. Model and calculation methods

The expressions for the total radiative capture cross-sections $\sigma(NJ, J_f)$ in the potential cluster model are given, for example, in works [24,25] and are written as

$$\sigma_c(NJ, J_f) = \frac{8\pi K e^2}{\hbar^2 q^3} \frac{\mu}{(2S_1+1)(2S_2+1)} \frac{J+1}{J[(2J+1)!!]^2} \times A_J^2(NJ, K) \sum_{L_i, J_i} P_J^2(NJ, J_f, J_i) I_J^2(J_f, J_i) \quad (2)$$

where for the electric orbital $EJ(L)$ transitions ($S_i = S_f = S$) we have [24]:

$$P_J^2(EJ, J_f, J_i) = \delta_{S_i S_f} \left[ (2J+1)(2L_i+1)(2J_i+1)(2J_f+1) \right] (L_i 0 J 0 | L_f 0)^2 \begin{Bmatrix} L_i & S & J_i \\ J_f & J & L_f \end{Bmatrix}^2,$$

$$A_J(EJ, K) = K^J \mu^J \left( \frac{Z_1}{m_1^J} + (-1)^J \frac{Z_2}{m_2^J} \right), \quad I_J(J_f, J_i) = \langle \chi_f | R^J | \chi_i \rangle. \quad (3)$$

Here, $q$ is the wave number of the initial channel particles; $L_f$, $L_i$, $J_f$, $J_i$, $S_f$, and $S_i$ are the angular moments of particles in the initial ($i$) and final ($f$) channels; $S_1$ and $S_2$ are spins of the initial channel particles; $m_1$, $m_2$, $Z_1$, and $Z_2$ are the masses and charges of the particles in the initial channel, respectively; $K$ and $J$ are the wave number and angular moment of the γ-quantum in the final channel; and $I_J$ is the integral over wave functions of the initial $\chi_i$ and final $\chi_f$ states as functions of relative cluster motion with the intercluster distance $R$. Emphasize that in our calculations here or earlier, we have never used such a notion as a spectroscopic factor (see, for example, [24]), i.e. its value is simply assumed to be equal to 1 [18-21,25].

For the spin part of the $M1(S)$ magnetic process in the used model the following expression is known ($S_i = S_f = S$, $L_i = L_f = L$) [24,25]

$$P_1^2(M1, J_f, J_i) = \delta_{S_i S_f} \delta_{L_i L_f} \left[ S(S+1)(2S+1)(2J_i+1)(2J_f+1) \right] \begin{Bmatrix} S & L & J_i \\ J_f & 1 & S \end{Bmatrix}^2,$$

$$A_1(M1, K) = \frac{e\hbar}{m_0 c} K \sqrt{3} \left[ \mu_1 \frac{m_2}{m} - \mu_2 \frac{m_1}{m} \right], \quad I_J(J_f, J_i) = \langle \chi_f | R^{J-1} | \chi_i \rangle, \quad J = 1. \quad (4)$$

Here, $m$ is mass of the nucleus, $\mu_1$, $\mu_2$ are the magnetic moments of clusters and the remaining symbols are the same as in the previous expression. The value of $\mu_n = -1.91304272\mu_0$ is used for the magnetic moment of neutron [26]. The correctness of the given above expression for the *M*1 transition is pre-tested on the basis of the radiative proton and neutron captures on $^7$Li and $^2$H at low energies in our previous studies [10,11,27-29].

To perform the calculations of the total cross section our computer program based on the finite-difference method (FDM) [30] has been rewritten. Now the absolute search precision of the binding energy for the n$^{16}$O system for different potentials is equal to $10^{-6}$–$10^{-8}$ MeV, the search precision of the determinant zero in the FDM, which determines the accuracy of the binding energy search, is equal to $10^{-14}$–$10^{-15}$, and the magnitude of the Wronskians of the Coulomb scattering wave functions is about $10^{-15}$–$10^{-20}$ [30].

The variational method (VM) with the expansion of the cluster wave function (WF) of the relative motion of the n$^{16}$O system on a non-orthogonal Gaussian basis [11,28] was used for additional control of the binding energy calculations: [21,30-33]

$$\Phi_L(r) = \frac{\chi_L(r)}{r} = Nr^L \sum_i C_i \exp(-\beta_i r^2), \qquad (5)$$

where $\beta_i$ and $C_i$ are the variational parameters and expansion coefficients, $N$ is the normalization coefficient of WF.

A computer program for solving of the generalized variational problem [30] has also been modified and used to control the accuracy of the FDM calculating of the binding energy and the wave function form. One of the possible and simple enough numerical algorithms for solving this problem with a simple program realization has been considered in [30,34].

The asymptotic constant of the ground state potential normally used to control the behavior of the wave function of the bound state at large distances, was calculated using the asymptotic behavior of the wave function in the Whittaker form [35]

$$\chi_L(r) = \sqrt{2k} C_W W_{-\eta_L + 1/2}(2kr), \qquad (6)$$

where $\chi$ is the numerical bound state wave function obtained from the solving of the Schrödinger equation and normalized to unity, *W* is the Whittaker function of the bound state, which determines the asymptotic behavior of the wave function and it is a solution of the same equation without the nuclear potential, i.e. the solution at large distances, *k* is the wave number caused by the binding energy of the channel, η is the Coulomb parameter defined hereinafter, *L* is the orbital angular momentum of the bound state.

The intercluster interaction potentials of the n$^{16}$O system, as usual, is chosen in simple Gaussian form [11,16]:

$$V(r) = -V_0 \exp(-\alpha r^2), \qquad (7)$$

where $V_0$ and $\alpha$ are the potential parameters usually obtained on the basis of description of the elastic scattering phase shifts at certain partial waves taking into account their resonance behavior or spectrum structure of resonance levels and, in the case of discrete spectrum, on the basis of description of the BS characteristics of the n$^{16}$O system. In both cases such potentials contain BS, which satisfy the classification of allowed states (AS) and FS according to Young tableaux given above.

In the present calculations the exact values of the neutron mass $m_n$ = 1.00866491597 amu [36], and the mass of $^{16}$O equals 15.994915 amu [37]. The value of the $\hbar^2/m_0$ constant is equal to 41.4686 MeV·fm$^2$. The Coulomb parameter $\eta = \mu \cdot Z_1 \cdot Z_2 e^2/(k \cdot \hbar^2)$ equals zero in this case, is represented as $\eta = 3.44476 \cdot 10^{-2} \cdot Z_1 \cdot Z_2 \cdot \mu/k$, where $k$ is the wave number in the fm$^{-1}$, defined by the interaction energy of the particles and $\mu$ is the reduced mass of the particles in amu.

### 3. Radiative neutron capture on $^{16}$O in cluster model

#### 3.1. *Classification of the orbital states*

Going to the analysis of the total cross sections of the $^{16}$O(n, γ)$^{17}$O capture, let's first consider the classification of the orbital states of the n$^{16}$O system according to Young tableaux. The GS of $^{16}$O corresponds to the Young tableau {4444} [16,18,38]. Recall that the possible orbital Young tableaux in the $N = n_1 + n_2$ particles system can be defined as the direct exterior product of the orbital tableaux of each subsystem, that gives {1} × {4444} → {5444} + {44441} for the n$^{16}$O system [39]. The first of these tableaux is compatible with the orbital angular moment $L = 0$ and it is forbidden, because five nucleons can not be in the *s*-shell, and the second tableau is allowed and compatible with the orbital angular moment equal to 1 [39].

Hence, in the potential of $^2S_{1/2}$ wave that corresponds to the first excited state of $^{17}$O in the n$^{16}$O channel and scattering states of these clusters, there is a forbidden state, and the $^3P$ scattering wave does not contain the forbidden state, but the allowed BS with {44441} tableau can be in both spectra – continuous and discrete. The GS of $^{17}$O in the n$^{16}$O channel, which has the energy of -4.1436 MeV [40], refers to the $^2D_{5/2}$ wave and also does not contain FSs. However, since we do not have the complete tables of products of Young tableaux for the system with the number of particles greater than eight [41] previously used for such calculations [18-20,27,42], then obtained above results should be considered only as a qualitative assessment of possible orbital symmetries in the bound states of $^{17}$O in the n$^{16}$O channel.

#### 3.2. *Phases and potentials*

To perform the calculation of radiative capture within the MPCM it is need to know the potentials of the n$^{16}$O elastic scattering in $^2S_{1/2}$, $^2P_{1/2}$, $^2P_{3/2}$, $^2D_{3/2}$ and $^2D_{5/2}$ waves and the interaction of the ground $^2D_{5/2}$ and the $^2S_{1/2}$ first excited states of $^{17}$O in the n$^{16}$O channel. The experimental data for the total cross sections of radiative capture, obtained in [43,44], are presented for the transition just to these bound states.

As already mentioned, the scattering potentials are constructed on the basis of the elastic scattering phases obtained at the energy $E > 1.1$ MeV in Refs. [45,46]. For the energy range 0.2–0.7 MeV there are the results of the phase analysis [47] based on the measurements of differential cross sections for n$^{16}$O elastic scattering [48] in the resonance region at 0.433 MeV [40]. Later, new experimental data [49] on the excitation function at energies from 0.5 MeV to 6.2 MeV were given in the EXFOR database [50,51]. These data [49] have not been used to date in the phase shift analysis in the $^2D_{3/2}$ resonance region at the energy of 1.0 MeV [40].

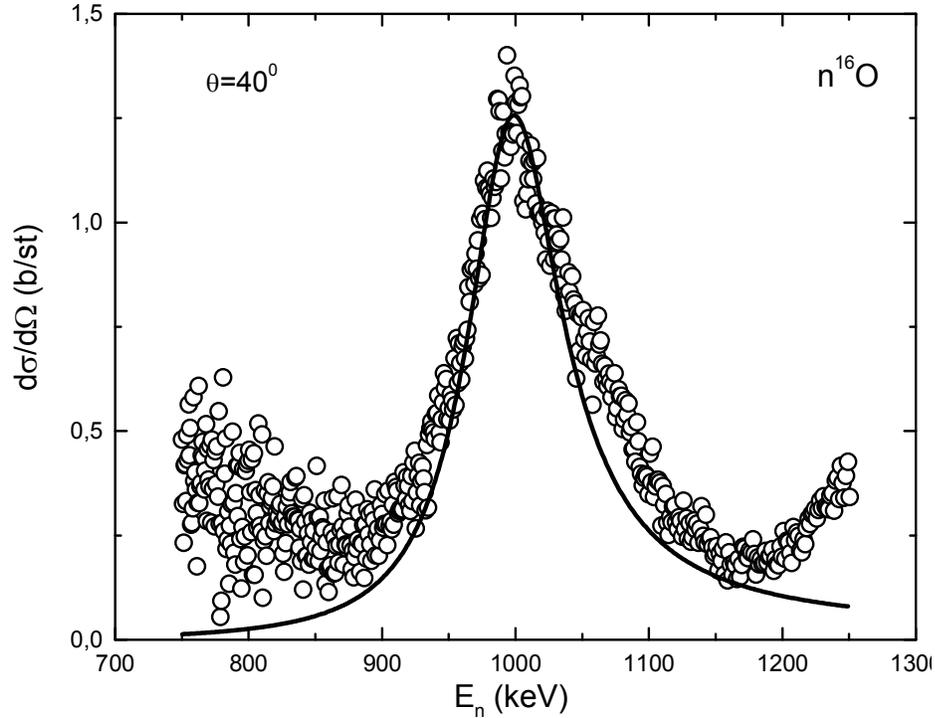

**Figure 1.** The excitation functions of the n$^{16}$O elastic scattering in the $^2D_{3/2}$ resonance region at 1.0 MeV. Data [49] are shown by the open circles. Solid line is the calculation of the cross sections with the potential given in the text.

Here, the data [49] were used to perform the phase shift analysis and for extraction the form of phase in the $^2D_{3/2}$ scattering wave. The used excitation functions at 40° in the laboratory system (l.s.) [49] are shown in figure 1 at the energy range from 0.75 MeV to 1.25 MeV (l.s.) by the open circles. The experimental errors which in some points reach 25% are not presented in figure 1, because they overload a lot the figure. This is caused by that in our analysis more than 500 points were used for the cross sections at the different energies from the excitation functions [49]. First it should be noted that below 0.7–0.8 MeV the ambiguity of the data [49] increases dramatically, however, to extract $^2D_{3/2}$ scattering phase shift is sufficient to consider the energy region shown in figure 1, which has a relatively small ambiguities and these data may be well used for carrying out the phase analysis. Earlier, we have performed the phase analysis in the similar systems n$^{12}$C [11,52], p$^{12}$C [53], p$^6$Li [54], and p$^{13}$C [55,56], mainly at the astrophysical energies.

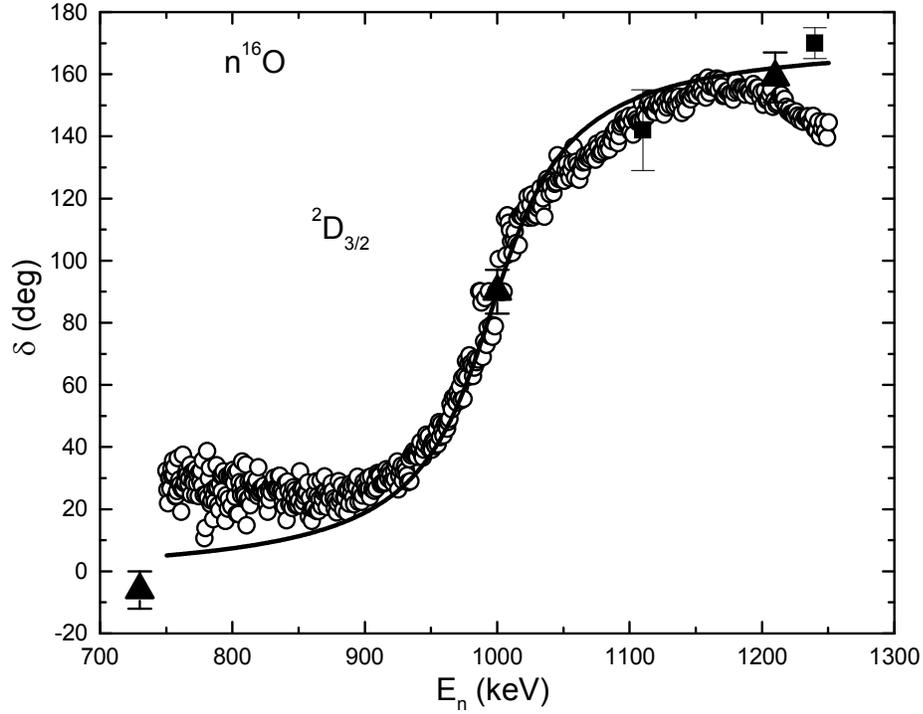

**Figure 2.** The $^2D_{3/2}$ phase shift of the n$^{16}$O elastic scattering at low energies. Open circles (○) are the results of our phase analysis carried out on the basis of the data [49]; black squares (■) show the results of phase analysis [45]; triangles (▲) correspond to the results of the analysis [47]; solid curve is the calculation of phase shift with the potential given in the text.

The details of the used method of searching the phase shifts in the elastic scattering of particles with spin 1/2$^+$0 are presented in Ref. [11], the main expressions are given in Refs. [11,17], and the results of the present analysis of the n$^{16}$O elastic scattering in the energy range from 0.75 MeV to 1.25 MeV are shown by the open circles in figure 2. Black squares in figure 2 shows the results of the phase analysis of [45] obtained at the energy $E > 1.1$ MeV, and the triangles are the results of the analysis of Ref. [47]. The quantity $\chi^2$ has a mean value of 4.7·10$^{-3}$ with a maximum value of the partial quantity $\chi^2_i = 0.6$ at the energy of 999.5 keV, since only one value of the cross sections is considered for each energy value. To describe the cross sections in the excitation functions [49], at least at energies up to 1.2–1.25 MeV, is not required to take into account the $^2S_{1/2}$ scattering phase, because its presence does not change the magnitude of $\chi^2$, i.e., its value can be assumed to be equal to zero.

For a description of the $^2D_{3/2}$ phase shift obtained from the phase analysis, one can use the potential of the form (7) without the FS with the parameters:

$$V_D = 95.797 \text{ MeV, and } \alpha_D = 0.17 \text{ fm}^{-2}, \qquad (8)$$

which leads to the resonance energy of 1000 keV at the phase shift of 90.0 (0.1)° with the level width of 88 keV (l.s.) or 83 keV in the center-of-mass system (c.m.). At the same time, in Table 17.17 of [40] the width is equal to 102 keV (l.s.) or 96 keV (c.m.) at the level energy of 1000±2 keV (l.s.). The energy dependence of the $^2D_{3/2}$ phase shift of potential (8) is shown in figure 2 by a solid curve. Such a potential describes well the behavior of the scattering phase shift in the resonance region and is consistent with the previous scattering phase shift extractions [45,47]. The shape of the cross sections in the excitation functions, calculated with the $^2D_{3/2}$ phase shift of potential (8) at zero values of other phases is shown in figure 1 by the solid curve. To estimate the width of the resonance was used expression (9).

$$\Gamma = 2(d\delta/dE)^{-1}, \qquad (9)$$

Next, we will consider the total radiative capture cross section with the $E1$ transitions from the $^2P_{3/2}$ resonance in n$^{16}$O scattering at 433 keV to the ground and first excited states of $^{17}$O. To build the $^2P_{3/2}$ scattering potential the data on the position and width of this level from the review [40] (see Table 17.17) and the results of the phase analysis [47] shown by triangles in figure 3 have been used. As a result, is found that to describe the resonant $^2P_{3/2}$ scattering phase shift at 433(2) keV (l.s.) with the width of 45 keV (c.m.) or 48 keV (l.s.) [40,48] the potential without the FS with the following parameters is needed:

$$V_P = 1583.545 \text{ MeV, and } \alpha_P = 6.0 \text{ fm}^{-2}, \qquad (10)$$

which leads to the level width equal to 44 keV (c.m.) or 47 keV (l.s.) at the resonance of 433 keV (l.s.), i.e., its phase at this energy is equal to 90.0(0.2)°, and the total dependence of the phase on the energy in the resonance region is shown in figure 3 by the solid curve.

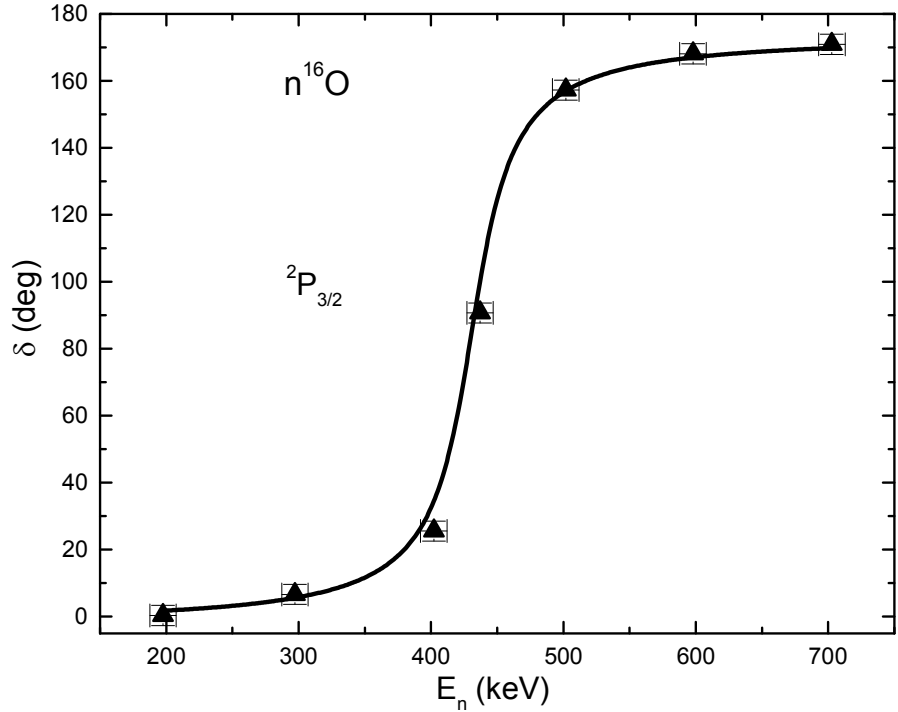

**Figure 3.** The $^2P_{3/2}$ phase shift of the n$^{16}$O elastic scattering at low energies. Triangles (▲) show the results of the phase shift analysis of Ref. [47]. The solid curve corresponds to the calculation of phase shift with the potential given in the text.

The $V_0 = 0$ MeV, i.e. zero phase shifts, were used for the potentials of the non-resonance $^2P_{1/2}$ and $^2D_{5/2}$ scattering waves, because in spectra of $^{17}$O below 1.0–1.5

MeV the levels with $J = 1/2^-$ and $5/2^+$ are not observed, and they have no FS. Next, one should note that the potential is built unambiguously on the basis of known energy of the resonance level in spectra of $^{17}$O (see Table 17.17 [40]) and its width, i.e., it is impossible to find other parameters $V$ and $\alpha$, which can properly reproduce the resonance energy of the level and its width if the number of the FS is given. In this case the number of the FS is equal to zero. The depth of this potential unambiguously determines the position of the resonance, i.e., the resonance energy of the level, and the level width sets a certain width of this resonance state.

Next to perform the calculations of the radiative capture within the MPCM the potentials of n$^{16}$O clusters interaction in the bound states are needed. The electromagnetic transitions to the ground bound state of $^{17}$O in the n$^{16}$O channel [40] with $J^\pi, T = 5/2^+, 1/2$ at the energy -4.1426 MeV and to the first excited state of this nucleus with $J^\pi = 1/2^+$ at -3.2729 MeV will be considered. The width of such potentials was fixed on the basis of the correct description of the binding energy and the charge radius of $^{17}$O that is equal to 2.710(15) fm [40], and then the comparison of the asymptotic constants of the n$^{16}$O channel with other data was performed.

As a result, for the $^2D_{5/2}$ potential of the GS of $^{17}$O in the n$^{16}$O channel without FS the following parameters were found:

$$V_{D0} = 102.2656782 \text{ MeV, and } \alpha_{D0} = 0.15 \text{ fm}^{-2}. \qquad (11)$$

These parameters allow to get the binding energy of -4.1436000 MeV with the accuracy of $10^{-7}$ MeV, the charge radius is equal to 2.71 fm, the mass radius of 2.73 fm and the asymptotic constant on the distance range 6–16 fm equal to $C_W = 0.75(1)$, defined by using the given above expression [35]. The charge radius of the neutron was equal to zero, its mass-radius was assumed to be equal to the proton radius of 0.8775 (51) fm [26], and the value of the charge radius of $^{16}$O was equal to 2.710(15) fm [57]. Methods of calculation of the energy, the wave functions and the radii are given in Refs. [10,11,18,19]. In Ref. [58] for the AC of the ground state the value of 0.9 fm$^{-1/2}$ was obtained, which after recalculation with the factor $\sqrt{2k} = 0.933$ to a dimensionless quantity gives the value 0.96. This recalculation is required, since in these works the different definition of AC was used, which is different from ours by a factor $\sqrt{2k}$. In [35] for this constant in dimensionless form the value 0.77(8) is given, which practically coincides with the value obtained above.

For the $^2S_{1/2}$ potential of the first excited state of $^{17}$O in the n$^{16}$O channel with one FS the following parameters were found

$$V_{S1} = 81.746753 \text{ MeV, and } \alpha_{S1} = 0.15 \text{ fm}^{-2}, \qquad (12)$$

parameters (12) lead to the binding energy of -3.2729000 MeV relative to the threshold of the n$^{16}$O channel of $^{17}$O with the accuracy of $10^{-7}$ MeV, the charge radius of 2.71 fm, the mass radius of 2.80 fm and AC on the distance range 6–17 fm is equal to $C_W = 3.09(1)$. The value 3.01 fm$^{-1/2}$ for the AC for this level was obtained in Ref. [58], that after the recalculation with $\sqrt{2k} = 0.88$ gives 3.42. As is seen, in this case the asymptotic constant value is in reasonable agreement with the other results.

To control the accuracy of the BS energy calculation a variational method [30] was used. This method for the GS on the grid with the dimension $N = 10$ and the independent variation of the parameters for the potential (11) allowed to obtain the energy of -4.1435998 MeV. The charge radius and the asymptotic constant in the interval of 6–16 fm do not differ from the values obtained in the FDM calculations. Since the variational energy is decreasing with the basis dimension increasing and gives the upper limit of the true binding energy, and the finite-difference energy is increasing with decreasing of the step size and increasing of the number of steps [10,11,18,21,30], so the real binding energy in this potential can take on the average value of -4.1435999(1) MeV. Thus, the accuracy of the binding energy calculation by two methods and two different computer programs is at the level of ±0.1 eV and in full compliance with the defined in the FDM program error of the binding energy search $10^{-7}$ MeV.

For the energy of the first excited state on the grid with the dimension $N = 10$ and the independent variation of the parameters for the potential (12) the energy of -3.2728998 MeV was obtained. The charge radius and the AC in the interval of 6–20 fm do not differ from the values obtained within the FDM. Here, the real binding energy can take on the average value of -3.2728999(1) MeV, i.e. the accuracy of the energy calculation by two ways and two different computer programs is also equal to ±0.1 eV.

### 3.3. *Total cross sections of the radiative capture*

Earlier, the radiative neutron capture reaction on $^{16}$O was examined basing on the model of direct capture in Ref. [59], where it was shown that it is possible to describe the available experimental data [43,44] in the range of 20–280 keV. Then basing on the folding model [60] was shown that one managed to describe the experimental data [43,44] in the energy range 20–60 keV. Furthermore, on the basis of the GCM, taking into account only the *E*1 transition in Ref. [61,62] the correct description of the total cross section in the energy range from 20 to 280 keV [43,44] have been obtained. Finally in Ref. [63] on the basis of the GCM and the microscopic R-matrix analysis based on the *E*1 and *M*1 processes on the whole one managed to reproduce correctly the experimental total cross sections at the energies from 25 meV [64-66] to 20–280 keV [43,44], i.e., to the resonance at 433 keV corresponding to the $^2P_{3/2}$ wave of n$^{16}$O-scattering [40], and to predict the possible behavior of the cross sections in the range of the $^2D_{3/2}$ resonance. However, the energies near the $^2P_{3/2}$ resonance, at which there is a comparatively new experimental data [67] from 160 to 560 keV, have not been considered till now.

In the present analysis based on the MPCM with the FS [10,11,18-21] by considering the data [67] the total cross sections of the neutron capture reaction on $^{16}$O in the energy range 10 meV–1.3 MeV will be considered. In this case one takes into account the *E*1 transitions:
1. from the resonance $^2P_{3/2}$ scattering wave to the $^2D_{5/2}$ ground bound state;
2. from the $^2P_{1/2}$ and $^2P_{3/2}$ scattering waves to the $^2S_{1/2}$ first excited state which is bound in the n$^{16}$O channel of $^{17}$O.

In addtion, the following M1 processes will be discussed:
3. from the resonance $^2D_{3/2}$ wave to the $^2D_{5/2}$ ground state of $^{17}$O;
4. from the $^2S_{1/2}$ scattering wave to the $^2S_{1/2}$ first excited state.

The total cross sections of the other possible transitions, for example, the *E*2

transition from the resonance $^2D_{3/2}$ wave to the $^2S_{1/2}$ first excited state or the $M1$ transition from the $^2D_{5/2}$ non-resonance scattering wave to the $^2D_{5/2}$ ground state and others, are on 2–3 orders less than ones mentioned above. Separate consideration of the transitions to the GS and the first excited state has been made possible by the measurements carried out in Ref. [43,44] in the energy range 20–280 keV. Moreover, the results of measurements [43,44] have shown that the cross sections of the capture to the first excited state is in 3–4 times greater than another one to the GS of $^{17}$O.

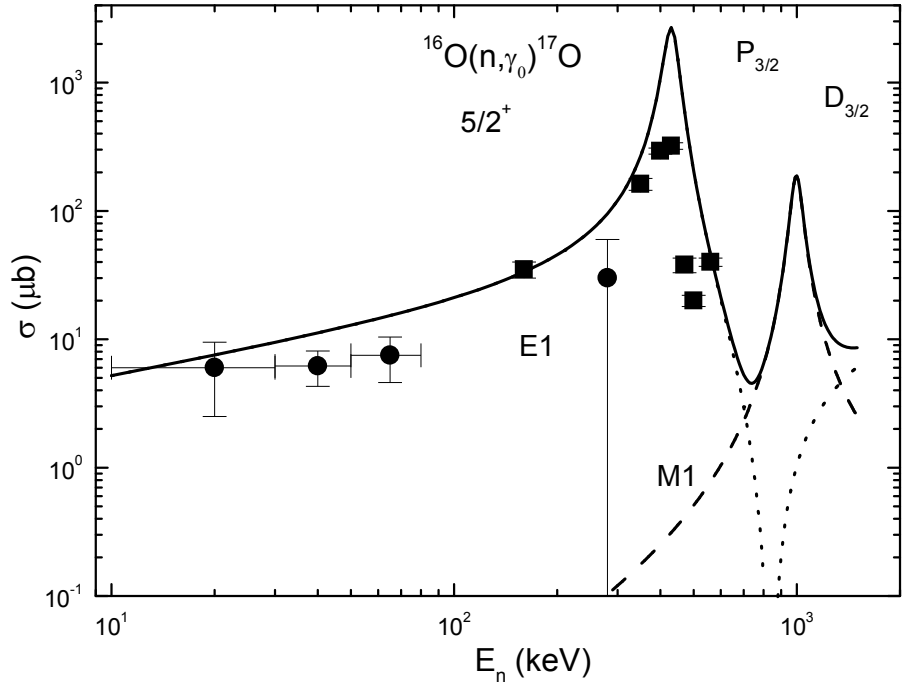

**Figure 4.** The total cross sections of the $E1$ radiative neutron capture on $^{16}$O to the $5/2^+$ ground state. The experimental data: ● – [43,44], ■ – [67]. Curves correspond to the present calculations with the given in the text potentials.

The results of our calculations of the total cross section for the $M1$ and $E1$ transitions to the GS, i.e., the processes 1 and 3 from the given above list, with the potentials (9-11) in comparison with experimental data [43,44,67] are presented in figure 4 by the solid curve. The cross sections of $M1$ transition from the $^2D_{3/2}$ scattering wave at energies up to 1.3 MeV, i.e., near the $^2D_{3/2}$ resonance, are represented by the dashed curve (process 3), and for the $E1$ process by the dotted curve (process 1). In figure 4 the results of measurements of the total cross sections, performed in the resonance region from 160 to 560 keV in Ref. [68]. In this Ref. [68], apparently, the summary data for the transitions to the GS and the first excited state are given. From figure 4 is seen that our calculations, performed with taking into account the $M1$ and $E1$ processes, reproduce acceptably the results of experimental measurements of the total cross sections of the transition to the GS of $^{17}$O [43,44], which gradually decreases with the energy decreasing. And it is necessary to note that the potentials of the $^2P_{3/2}$ and $^2D_{3/2}$ scattering waves and the $^2D_{5/2}$ bound state of the n$^{16}$O cluster system, which does not contain the FS, were constructed on the basis of the simple assumptions about the consistency of the scattering potential with the scattering phases, and about the consistency of the BS potential with the main characteristics of the GS of $^{17}$O (the binding energy, charge radius, and AC).

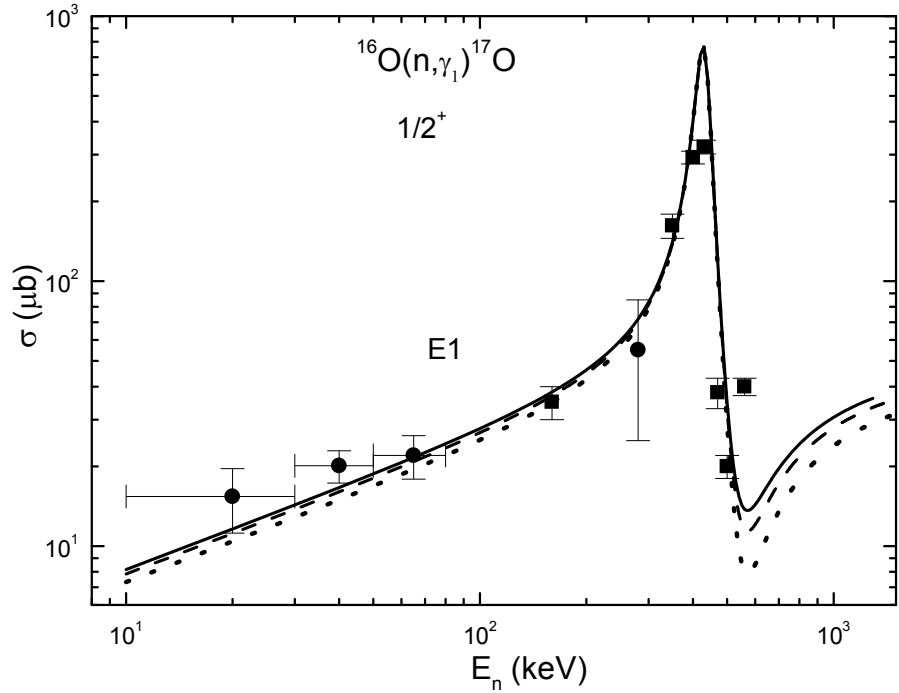

**Figure 5.** The total cross sections of the $E1$ radiative neutron capture on $^{16}$O to the $1/2^+$ first excited state of $^{17}$O. Experimental data: ● – [43,44], ■ – [68]. Curves are our calculation of the total cross section for the potentials given in the text.

The results of calculation of the cross sections for the $E1$ transitions from the $^2P_{3/2}$ and $^2P_{1/2}$ scattering waves to the first excited state of $^{17}$O (process 2) are shown in figure 5 by the solid curve. This calculation was performed in the energy range from 10 keV to 1.3 MeV without the second resonance in the $^2P_{3/2}$ wave at 1312 keV [40], i.e., for the $^2P_{1/2} + ^2P_{3/2} \to ^2S_{1/2}$ process. Here the dotted curve shows the results of measurements of the cross sections from Ref. [43,44] for the transition to the $1/2^+$ first excited level in the energy range 20–280 keV. The squares show the measurements of the total cross sections [67]. It seems that the measurement of [67] are in better agreement with the earlier experimental results for the transitions to the first excited state [43,44]. In this case, they are quite reasonably described in our calculations at energies 20–560 keV. For the potential of the non-resonance $^2P_{1/2}$ wave without the FS the zero depth was used, and for the potential of the resonance $^2P_{3/2}$ wave interaction (10) was used.

By analogy with the earlier considered n$^{12}$C system [68], one can assume that the second excited state with $J^\pi = 1/2^-$ at 3.055 MeV relative to the GS [40] may belong to the $^2P_{1/2}$ wave. Then one should accept the presence of the bound AS in this partial wave which as usual must lead to a zero scattering phase shift. In this case, for example, the parameters of the $^2P_{1/2}$ potential are accepted a little deeper than they were determined for the interaction with $J^\pi = 3/2^-$, (see equation (10))

$$V_{1/2} = 1593.435350 \text{ MeV, and } \alpha_S = 6.0 \text{ fm}^{-2}. \tag{13}$$

It gives the approximate to zero phase shifts in the region up to 1 MeV, and leads to the binding energy of -1.08824 MeV relative to the n$^{16}$O threshold at the accuracy of the FDM of $10^{-5}$ MeV. The charge radius of $^{17}$O in the 1/2⁻ second excited state is equal to 2.70 fm, the mass radius of 2.65 fm, and the AC is equal to 0.22 in at the interval of 2–18 fm. The results obtained for the total cross sections of the capture for the $E$1-process $^2P_{1/2} + ^2P_{3/2} \to\ ^2S_{1/2}$ with such potential and potential (10) almost coincide with the previous ones; they are shown in figure 5 by the dashed curve.

It is not succeed to find the values of the AC for the second excited state at 3.055 MeV with $J^\pi = 1/2^-$, so it is not possible to compare the obtained above value of the AC. In order to, at least partially, get rid of the existing ambiguity of the parameters of this scattering potential let's consider another version of it with a wider interaction well

$$V_{1/2} = 270.711230 \text{ MeV, and } \alpha_S = 1.0 \text{ fm}^{-2}. \quad (14)$$

This potential also leads to a near-zero phase shifts, the binding energy of -1.08824 MeV with the accuracy of the FDM $10^{-5}$ MeV, the charge radius of 2.79 fm, the mass radius of 2.69 fm and the AC equal to 0.39 in the interval of 3–23 fm, which is nearly twice as much as for the previous potential. The capture total cross sections are shown in figure 5 by the dotted curve.

As is seen from these results the cross section does not strongly depend on the presence of the bound allowed state in the 1/2⁻ wave, if the width of such potential is comparable with the width of interaction in the $^2P_{3/2}$ wave and is in the region 1.0–6.0 fm, and the values of the AC are in the interval from 0.2 to 0.4. Hence, one cannot use this transition for the unambiguous choice of the form of the interaction potential in the $^2P_{1/2}$ scattering wave and determine the presence of the allowed bound state at $J = 1/2^-$ in it.

The total cross section for all transitions to the 5/2⁺ ground and the 1/2⁺ first excited states are shown in figure 6. It is evident that the results of the calculations quite reproduce the data [43,44], and the measurement results [67] are mainly below the calculated curve. Note that here the possible at the lowest energies $M$1 transitions have not been considered yet, but these transitions can increase somewhat the total cross section at the minimum energy shown in figure 6.

In this way, the considered above methods of constructing of the clusters interaction potentials allow, on the whole, to reproduce correctly experimental data for the total cross sections of the radiative capture at energies from 20 keV to 560 keV. However, in the case of the $M$1 transition to the first excited state in this cluster system the used above criteria are not sufficient for the unambiguous definition of the potential, and, as will be seen below, it is necessary to vary its parameters to describe better the experimental data [64-66] at the lowest energy 25 meV.

Going to the low-energy region, note that below 100 eV the capture cross section is completely determined by the $M$1 process with the transition from the $^2S_{1/2}$ scattering wave to the first excited state of $^{17}$O nucleus (process 4 in the given above list of transitions). The potential of the $^2S_{1/2}$ wave of the n$^{16}$O scattering contains the forbidden state, as follows from the considered above classification of cluster states. This potential at the energies under consideration must lead to a zero phase shift, but as it has the forbidden state, its depth cannot be equal to zero. The form of this potential was refined in order to describe correctly the cross section at 25 meV. The parameters of this potential are

$$V_S = 10.0 \text{ MeV, and } \alpha_S = 0.03 \text{ fm}^{-2}. \quad (15)$$

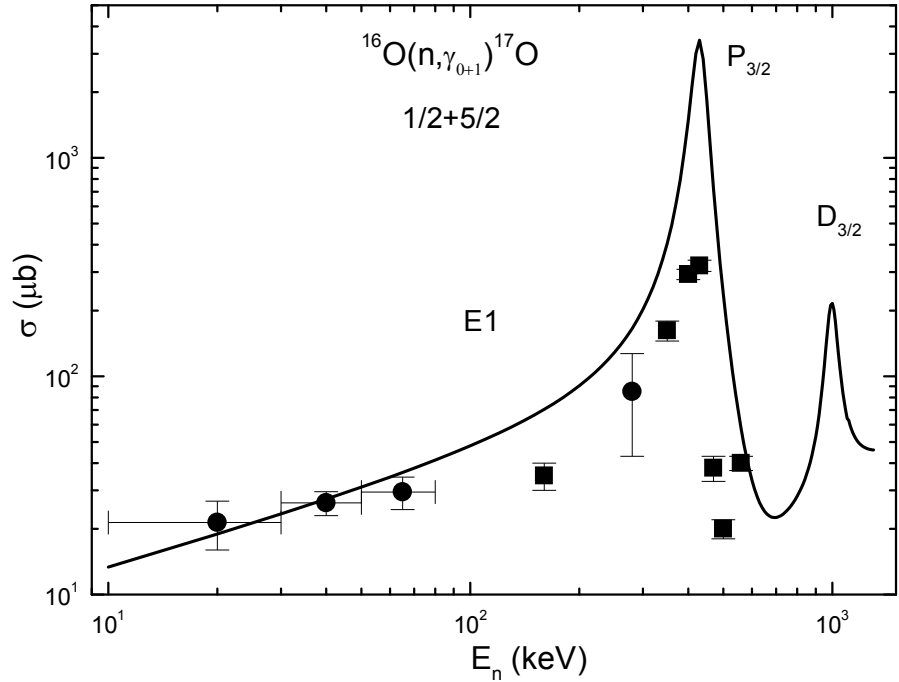

**Figure 6.** The total cross sections of the $E1$ radiative neutron capture on $^{16}$O to the $5/2^+$ ground and the $1/2^+$ first excited state of $^{17}$O. Experimental data: ● – [43,44], ■ – [67]. The continuous curve – our calculation of the total cross section.

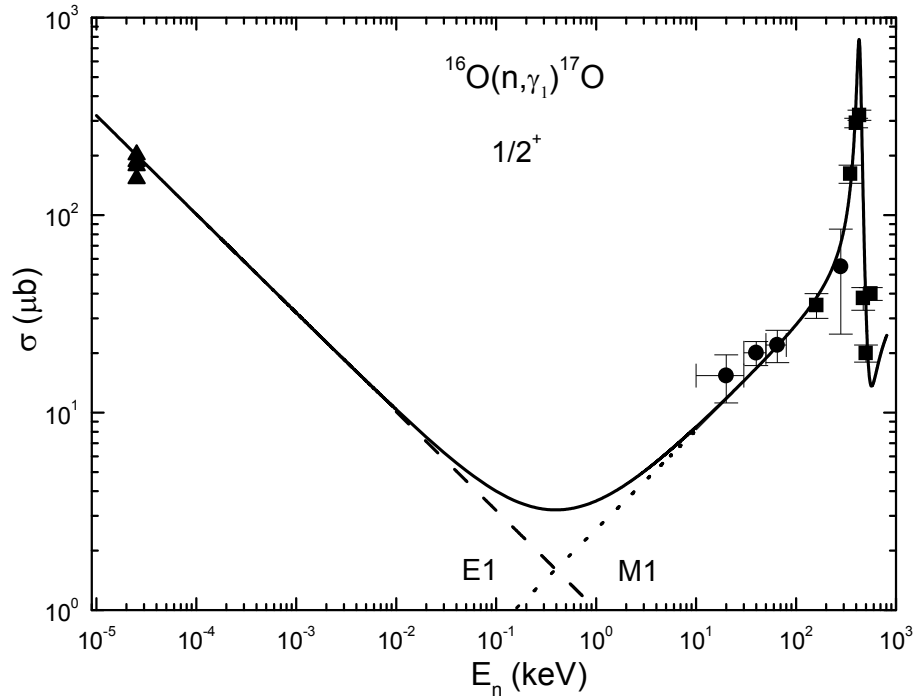

**Figure 7.** The total cross sections of $E1$ and $M1$ neutron radiative capture on $^{16}$O on the first excited $1/2^+$ state of $^{17}$O. Experimental data: ● – [43,44], ■ – [67], ▲ – [64-66]. The solid curve is our calculation of the total cross section for the potentials given in the text.

The results of calculations of the total cross sections with the $M1$ transition at 10 meV–1.0 MeV are shown in figure 7 by the solid curve. At the energy 25 meV the triangles show the results of experimental measurements [64-66], which are in the range 150–200 µb. The dashed curve in figure 7 corresponds to the results for the cross sections caused only by the $M1$ process, and the $E1$ transitions are shown in this figure by the dotted curve. As is seen from this figure the $E1$ cross section drops dramatically and at 100 eV is about three times smaller than the cross section of the $M1$ transition.

An additional point to emphasize is that only $^2S_{1/2}$ scattering potential with the FS allows one to describe correctly the capture total cross sections at the energy of 25 meV. If the potential without the FS is used, none of its parameters cannot reproduce correctly the behavior of the total cross sections at this energy. Namely, at the energy 25 MeV for potential (15) $^2S_{1/2}$ scattering phase is equal to 0.00812° (in case of using the newer value of constant $\hbar^2/m_0 = 41.8016$ MeV·fm$^2$ the phase shift is equal to 0.00692°) and capture cross section takes a value 202 µb. In case of using of the $^2S_{1/2}$ potential without the FS and with the parameter $V_0 = 0$ MeV, its phase shift equals zero, and the cross section at 25 meV increases up to 12.5 mb. Another variant of the $^2S_{1/2}$ potential without the FS, for example, with the parameters $V_0 = 3.18$ MeV and $\alpha = 0.1$ fm$^{-2}$, gives the same scattering phase 0.00812° and leads to further increasing of the cross section, i.e. to the value 37.9 mb. Thereby, the description of the cross section at the energy 25 meV within the MPCM methods is possible only by using the $^2S_{1/2}$ scattering potential containing the forbidden state.

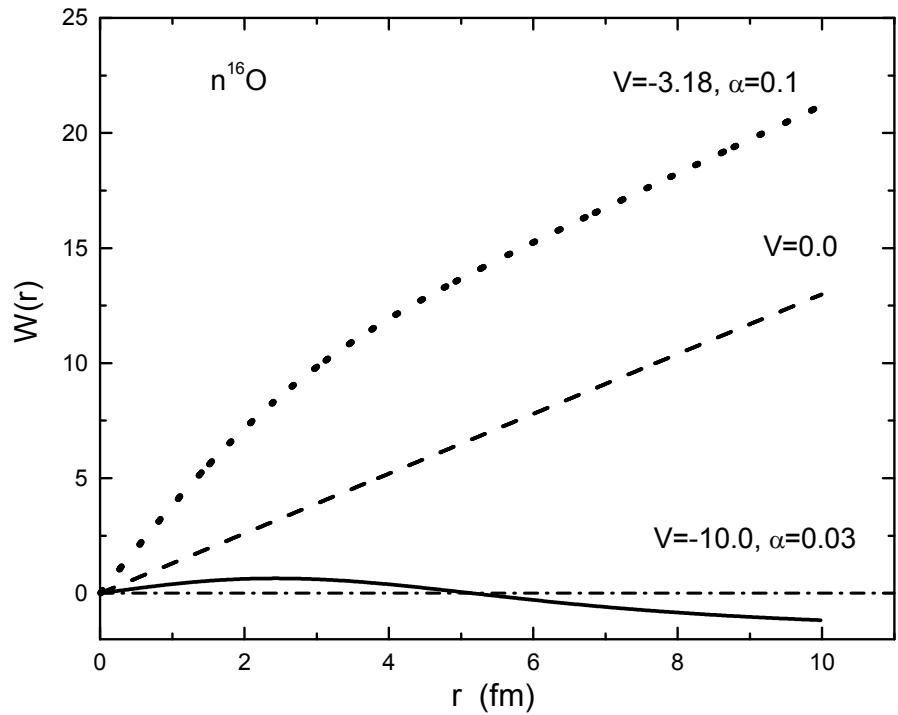

**Figure 8.** Wave functions of the scattering potentials at 25 meV

Such a great difference in the results may be explained by behavior of the scattering wave function of these potentials at 25 meV, as shown in figure 8. For the potential without the FS and with $V_0 = 0$ at 25 meV it is the straight line (the dashed curve), with a depth -3.18 MeV it is the dotted line, and for interaction (15) it oscillates even at such a low energy and has a node at 5.17 fm (the solid curve in figure 8).

Since at the energies from 10 meV to, approximately, 10 eV the calculated cross-section shown in figure 7 by the solid line is almost a straight line, it can be approximated by a simple function of the form:

$$\sigma_{ap}(\mu b) = \frac{A}{\sqrt{E_n(keV)}}. \qquad (16)$$

The value of constant $A = 1.0362$ µb·keV$^{1/2}$ was determined from a single point of the cross-sections with a minimal energy of $10^{-5}$ keV. Furthermore, it is possible to consider the absolute value of the relative deviation of the calculated theoretical cross-sections and the approximation of the cross-section (17) by this function given above in the energy range from 10 meV to 10 eV.

$$M(E) = \left|[\sigma_{ap}(E) - \sigma_{theor}(E)]/\sigma_{theor}(E)\right|. \qquad (17)$$

At the energies below 10 eV, the relative deviation is equal to 2.5%, and up to 1 eV does not exceed 0.5%. It may be assumed that this form of energy dependence of the total cross section will be stored at lower energies, and the cross section, for example, at the energy 1 µeV, takes the value 32.8 mb.

## 4. Conclusion

In this way, the quite acceptable results for description of the experimental total cross sections of the radiative neutron capture on $^{16}$O at the energies from 25 meV to 560 keV have been obtained for the considered intercluster potentials of the n$^{16}$O interaction, which answer to the classification of the states according to Young tableaux.

The potential of the $^2S_{1/2}$ scattering wave with the FS, the phase of which is close to zero, have been obtained. This potential allows us to describe correctly the behavior of the experimental cross sections at the lowest energies. It was shown that the description of the low-energy cross sections is possible only if there is a forbidden state in such potential. Interaction of the $^2S_{1/2}$ scattering wave without the FS, in principle, does not allow us to reproduce correctly the value of the capture cross section at such a low energy.

In the future it is necessary to carry out new experimental measurements of the total capture cross sections at 1 eV–1 keV, where the calculations predict the well-defined behavior of the cross-sections with a smooth minimum at the energy 0.4 keV and with the value 3 µb (figure 7). In addition, near 1.0 MeV, i.e., in the region of the $^2D_{3/2}$ resonance, a well-defined value of the second maximum of the cross sections (figure 4) is also received. In both cases obtained results are somewhat different from the similar data obtained earlier in [63] and, apparently, only the new experimental measurements are able to eliminate this discrepancy.

In conclusion note that there are twenty one cluster system considered on the basis of the modified potential cluster model with the classification of the orbital states according to the Young tableaux, in which it is possible to obtain acceptable results on the description of the characteristics of the processes of the radiative capture of nucleons or light clusters, mainly, on the 1p shell nuclei. The properties of these cluster nuclei, some characteristics and the considered cluster channels are shown in Table 1. Recent results are presented in [42,68-77].

**Table 1.** The characteristics of nuclei and cluster systems, and references to works in which they were considered.[+)]

| No. | Nucleus ($J^\pi$,T) | Cluster channel | $T_z$ | $T$ | Reference |
|---|---|---|---|---|---|
| 1. | $^3$H (1/2$^+$,1/2) | n$^2$H | -1/2 + 0 = -1/2 | 1/2 | [10,18,28] |
| 2. | $^3$He (1/2$^+$,1/2) | p$^2$H | +1/2 + 0 = +1/2 | 1/2 | [11,28,42] |
| 3. | $^4$He (0$^+$,0) | p$^3$H | +1/2 - 1/2 = 0 | 0 + 1 | [10,18,19] |
| 4. | $^6$Li (1$^+$,0) | $^2$H$^4$He | 0 + 0 = 0 | 0 | [10,18,19] |
| 5. | $^7$Li (3/2$^-$,1/2) | $^3$H$^4$He | -1/2 + 0 = -1/2 | 1/2 | [10,18,19] |
| 6. | $^7$Be (3/2$^-$,1/2) | $^3$He$^4$He | +1/2 + 0 = +1/2 | 1/2 | [10,18,19] |
| 7. | $^7$Be (3/2$^-$,1/2) | p$^6$Li | +1/2 + 0 = +1/2 | 1/2 | [10,20,69] |
| 8. | $^7$Li (3/2$^-$,1/2) | n$^6$Li | -1/2 + 0 = -1/2 | 1/2 | [11,68,70] |
| 9. | $^8$Be (0$^+$,0) | p$^7$Li | +1/2 - 1/2 = 0 | 0 + 1 | [10,19,20] |
| 10. | $^8$Li (2$^+$,1) | n$^7$Li | -1/2 - 1/2 = -1 | 1 | [11,71,76] |
| 11. | $^{10}$B (3$^+$,0) | p$^9$Be | +1/2 - 1/2 = 0 | 0 + 1 | [10,18,19] |
| 12. | $^{10}$Be (0$^+$,1) | n$^9$Be | -1/2 - 1/2 = -1 | 1 | [74] |
| 13. | $^{13}$N (1/2$^-$,1/2) | p$^{12}$C | +1/2 + 0 = +1/2 | 1/2 | [10,20] |
| 14. | $^{13}$C (1/2$^-$,1/2) | n$^{12}$C | -1/2 + 0 = -1/2 | 1/2 | [11,68,77] |
| 15. | $^{14}$N (1$^+$,0) | p$^{13}$C | +1/2 - 1/2 = 0 | 0 + 1 | [56,75] |
| 16. | $^{14}$C (0$^+$,1) | n$^{13}$C | -1/2 - 1/2 = -1 | 1 | [11,68,77] |
| 17. | $^{15}$C (1/2$^+$,3/2) | n$^{14}$C | -1/2 – 1 = -3/2 | 3/2 | [11,71] |
| 18. | $^{15}$N (1/2$^-$,1/2) | n$^{14}$N | -1/2 + 0 = -1/2 | 1/2 | [11,71] |
| 19. | $^{16}$N (2$^-$,1) | n$^{15}$N | -1/2 - 1/2 = -1 | 1 | [11,72] |
| 20. | $^{16}$O (0$^+$,0) | $^4$He$^{12}$C | 0 + 0 = 0 | 0 | [19,73] |
| 21. | $^{17}$O (5/2$^+$,1/2) | n$^{16}$O | -1/2 + 0 = -1/2 | 1/2 | Present work |

[+)] $T$ – isospin and $T_z$ – its projection, $J^\pi$ – total moment and parity.

**Acknowledgments**

The work was performed under the grant No. 0151/GF2 "Studying of the thermonuclear processes in the primordial nucleosynthesis of the Universe" of the Ministry of Education and Science of the Republic of Kazakhstan.
   In conclusion, the authors express their deep gratitude to Blokhintsev L.D., Burkova N.A., Mukhamedzhanov A., Yarmukhamedov R. for extremely useful discussion of the certain parts of this work.